\documentclass[journal]{IEEEtran}
\IEEEoverridecommandlockouts
\usepackage[top=0.75in, bottom=1 in, left=0.625in, right=0.625in]{geometry}

\usepackage{amsfonts}
\usepackage{amsmath}
\usepackage{amssymb}
\usepackage{pifont}
\usepackage{urwchancal}
\usepackage{algorithm}
\usepackage{algorithmicx}
\usepackage{algpseudocode}
\usepackage{dsfont}
\usepackage{authblk}
\usepackage{bbm}
\usepackage{graphicx}
\usepackage{epstopdf}
\usepackage{subfigure}
\usepackage{stfloats}
\usepackage{eqnarray}
\usepackage{makecell}
\usepackage{multirow}
\usepackage{enumerate}
\usepackage{booktabs}
\usepackage{cite}
\usepackage{balance}
\usepackage{color}
\usepackage{bm}
\usepackage{caption}
\usepackage{mathtools}

\newcommand{\cn}{{\cal CN}}

\newcommand{\inter}{\mbox{\tiny INTER}}
\newcommand{\intra}{\mbox{\tiny INTRA}}

\newcommand{\req}{\mbox{\tiny REQ}}
\newcommand{\st}{\mbox{s.t.\;}}

\newcommand{\sinr}{\mbox{SINR}}

\newcommand{\tr}{\mbox{Tr}}

\newcommand{\hh}{\mbox{\tiny H}}

\newcommand{\A}{\mbox{\tiny A}}
\newcommand{\B}{\mbox{\tiny B}}
\newcommand{\D}{\mbox{\tiny D}}

\newcommand{\F}{\mbox{\tiny F}}

\newcommand{\Y}{{\cal Y}}

\newcommand{\W}{\bm W}
\newcommand{\w}{\bm w}

\newcommand{\h}{\bm h}

\DeclarePairedDelimiter\autobracket{(}{)}

\newcommand{\br}[1]{\autobracket*{#1}}

\newcommand{\cb}[1]{\left\{#1\right\}}
\newcommand{\abs}[1]{\left|#1\right|}
\newcommand{\norm}[1]{\left\|#1\right\|_{\F}}

\newcommand{\summ}{\sum\limits_{m=1}^M}
\newcommand{\sumn}{\sum\limits_{n=1}^{N_m}}

\newtheorem{proposition}{\textbf{Proposition}}

\begin{document}
\title{\Large Joint Precoding and Power Control in Small-Cell Networks With Proportional-Rate MISO-BC Backhaul}
\author[$\ddag$]{\mbox{Yanjie Dong},~\IEEEmembership{Student Member, IEEE}}
\author[$\dag$]{\mbox{Md. Jahangir Hossain},~\IEEEmembership{Senior Member, IEEE}}
\author[$\dag$]{\mbox{Julian Cheng,~\IEEEmembership{Senior Member, IEEE}}}
\author[$\ddag\S$]{\mbox{Victor C. M. Leung},~\IEEEmembership{Fellow, IEEE}}

\affil[$\ddag$]{Department of Electrical and Computer Engineering\\
The University of British Columbia\\
Vancouver\\
\mbox{BC}\\
Canada}
\affil[$\dag$]{School of Engineering\\
The University of British Columbia\\
Kelowna\\
BC\\
Canada}
\affil[$\S$]{College of Computer Science and Software Engineering\\
Shenzhen University\\
Shenzhen\\
China
\authorcr Emails: \{ydong16, vleung\}@ece.ubc.ca\\
\{julian.cheng, jahangir.hossain\}@ubc.ca
\thanks{
This work was supported in part by the National Natural Science Foundation of China under Grant 61671088, in part by a UBC Four-Year Doctoral Fellowship, in part by the Natural Science and Engineering Research Council of Canada, and in part by the National Engineering Laboratory for Big Data System Computing Technology at Shenzhen University, China.}}

\maketitle
\begin{abstract}
In the small-cell networks with multiple-input-single-output broadcasting (MISO-BC) backhauls, the joint dirty-paper coding and power control are investigated for the \mbox{MISO-BC} backhauls and access links in order to minimize the system transmit power.
Considering the proportional rates of MISO-BC backhauls and flow-conservation constraints, the formulated optimization problem is \mbox{non-convex}.
Moreover, the formulated problem couples the precoding vectors with the power-control variables.
In order to handle the \mbox{non-convex} optimization problem and decouple the backhaul and access links, the structure of the formulated problem is investigated such that the optimal precoding vectors and optimal power-control variables are independently obtained.
Moreover, the optimal precoding vectors are obtained in closed-form expressions.
Simulation results are used to show the performance improvement over the benchmark scheme.
\end{abstract}
\pagestyle{empty}
\thispagestyle{empty}


\section{Introduction}
The volume of mobile data experiences a high-pace growth during the last decade under the development of mobile internet.
In order to support the ever-increasing data volume, the \mbox{small-cell} networks (SCNs) are proposed to offload the mobile data from the traditional infrastructure to the \mbox{small-cell} base stations (ScBSs).
In the SCN, the low-power ScBSs will be ultra-densely deployed to improve the spectrum efficiency of traditional infrastructure \cite{LiuSept.2018}.
Moreover, the ScBSs will improve the energy efficiency by reducing the distance between the transmitters and receivers \cite{LiuSept.2018}.
However, the successful application of SCNs depends on the reliable and \mbox{economical-friendly} backhauls, which connect the ScBSs to the core network.
Therefore, the backhauling technology has been considered as a key component for the fifth generation (5G) cellular networks \cite{SiddiqueOct.2015, ZhangDec.2016, DongDec.2017, XiangFeb.2018, MaMar.2018a, Dongtobepublished2019}.

The backhauling technology can be classified into two categories:  wired backhauls \cite{DongDec.2017, XiangFeb.2018} and wireless backhauls \cite{SiddiqueOct.2015, ZhangDec.2016}.
For example, the ScBSs can connect to the core network via the wired backhauls, such as optical fibers or digital subscriber lines \cite{DongDec.2017, XiangFeb.2018}.
Though the wired backhauls can provide \mbox{high-speed} connections, the capital expenditure is high.
In some scenarios (e.g., temporary usage of backhauls in the exhibitions and music concerts), the deployment of wired backhauls is not necessary or not available.
Therefore, the wireless backhauls are proposed as the promising alternatives to the wired backhauls due to the low cost and fast deployment \cite{SiddiqueOct.2015, ZhangDec.2016}.
Moreover, the wireless backhauls also extend the coverage of cellular networks to the remote area by providing the ScBSs with the plug-and-play backhauls.
Based on the spectrum bands, the wireless backhauls are categorized into millimeter-wave backhauls \cite{SiddiqueOct.2015, ZhangDec.2016}, optical wireless backhauls \cite{Hassantobepublished}, radio-frequency (RF) backhauls \cite{ZhaoOct.2015, VuMar.2016, YangJuly2018, ZhangApr.2018, ChenDec.2016, ZhaoMay2018, NguyenOct.2018, MostafaDec.2018} and hybrid backhauls \cite{MostafaDec.2018, HaoJan.2018}.
Several researchers have investigated the  characteristics of different spectrum bands for the wireless backhauls \cite{SiddiqueOct.2015, ZhangDec.2016}.

Due to the low-cost RF devices, we investigate the resource allocation in the SCNs with RF backhauls.
In the SCNs with RF backhauls, the current resource allocation algorithms mainly focus on the the admission control \cite{ZhaoOct.2015}, power minimization \cite{VuMar.2016, YangJuly2018}, system energy efficiency maximization \cite{ChenDec.2016, ZhangApr.2018}, system security \cite{ZhaoMay2018} and system capacity maximization \cite{NguyenOct.2018, HaoJan.2018, MostafaDec.2018}.
For example, Vu \emph{et al.}  proposed a power-control algorithm to minimize the transmit power of the RF backhauls subject to the communication quality-of-service (QoS) for the single-input-single-output broadcasting \mbox{(SISO-BC)} backhauls \cite{VuMar.2016}.
Since the proposed algorithm focuses on the power control of backhauls, it induces a suboptimal performance when the radio resources of backhauls and access links are considered.
Using the SISO-BC backhauls, Zhang \emph{et al.}  investigated the joint backhaul bandwidth allocation and access-link power control to maximize the system energy efficiency \cite{ZhangApr.2018}.
Since the multiple-input-single-output broadcasting (MISO-BC) channels have larger capacity region than the SISO-BC channels with the same pathloss, the linear precoding schemes were investigated to the joint resource allocation of backhauls and access links in the SCNs with MISO-BC backhauls \cite{ChenDec.2016, NguyenOct.2018, HaoJan.2018, MostafaDec.2018, ZhaoMay2018}.

Compared with the linear precoding schemes, the \mbox{dirty-paper} coding scheme (DPCS) can improve the rate of MISO-BC backhauls \cite{WangMay2015}.
When the DPCS is used by the MISO-BC backhauls, the impact of joint backhaul precoding and access-link power control has not been reported in the current literature.
Moreover, the current algorithms based on linear precoding scheme in \cite{ChenDec.2016, NguyenOct.2018, HaoJan.2018, MostafaDec.2018, ZhaoMay2018} cannot be used to design the precoding vectors and power-control variables for the MISO-BC backhauls and access links, respectively.
Besides, the fairness issue among the ScBSs is not considered in the SCNs with MISO-BC backhauls \cite{ChenDec.2016, NguyenOct.2018, HaoJan.2018, MostafaDec.2018, ZhaoMay2018}.
Therefore, we are motivated to investigate the joint precoding and power control (JPPc) problem in the SCNs with proportional-rate MISO-BC backhauls.
Our contributions are summarized as follows.
\begin{itemize}
  \item In order to minimize the system transmit power, we formulate the JPPc problem when the MISO-BC backhauls are used in the SCNs.
      We consider that the rates of backhauls satisfy a set of ratios in order to satisfy the fairness of ScBSs.
  \item The formulated JPPc problem is non-convex.
   Moreover, the precoding vectors are coupled with the power-control variables.
    Hence, the JPPc problem is challenging to solve via the standard solutions.
   \item Based on the structure of JPPc problem, we obtain a set of optimal rate ratios for the MISO-BC backhaul.
       Based on the optimal ratios, we can equivalently decouple the JPPc problem into two subproblems: access-link power-control subproblem and backhaul precoding subproblem.
      Therefore, the access-link power-control subproblem can be solved via the convex optimization toolbox.
      Different from \cite{SchubertJan.2004, WangMay2015}, we obtain the closed-form precoding vectors for the backhaul precoding subproblem when the encoding sequence of DPCS is given.
\end{itemize}

Notation: $\W^{\hh}$ denotes the hermitian of matrix $\W$.
$\mathbb{C}$ denotes the domain of complex values.
The expectation of a random variable is denoted as  $\mathds{E}\cb{\cdot}$.
The operator $\abs{\cdot}$ and $\norm{\cdot}$ respectively denote the determinant and Frobenius norm of a matrix.


The remaining of this paper is organized as follows.
In Section II, the system model and problem formulation are described.
In Section III, the optimal algorithm is proposed to obtain the optimal power-control variables and the closed-form precoding vectors.
Numerical results are presented in Section IV.
Section V concludes this paper.


\section{System Model and Problem Formulation}
\subsection{Overall System Description}
\begin{figure}[htb]
  \centering
  \includegraphics[width=3 in]{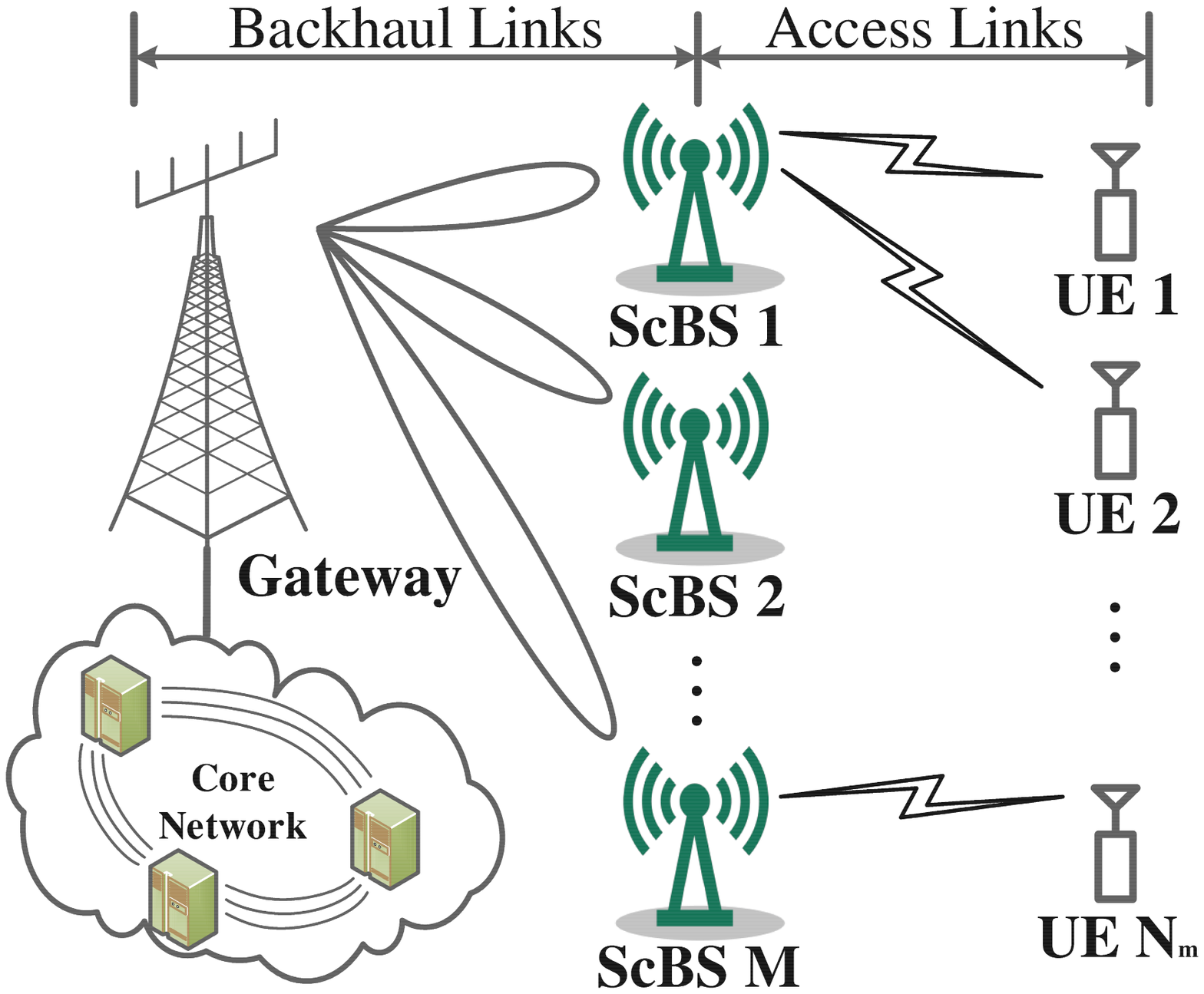}
  \vspace{0.3 cm}
  \caption{An illustration of the SCN with MISO-BC backhauls and RF access links.}\label{fg:001}
      \vspace{0.2 cm}
  \includegraphics[width=3 in]{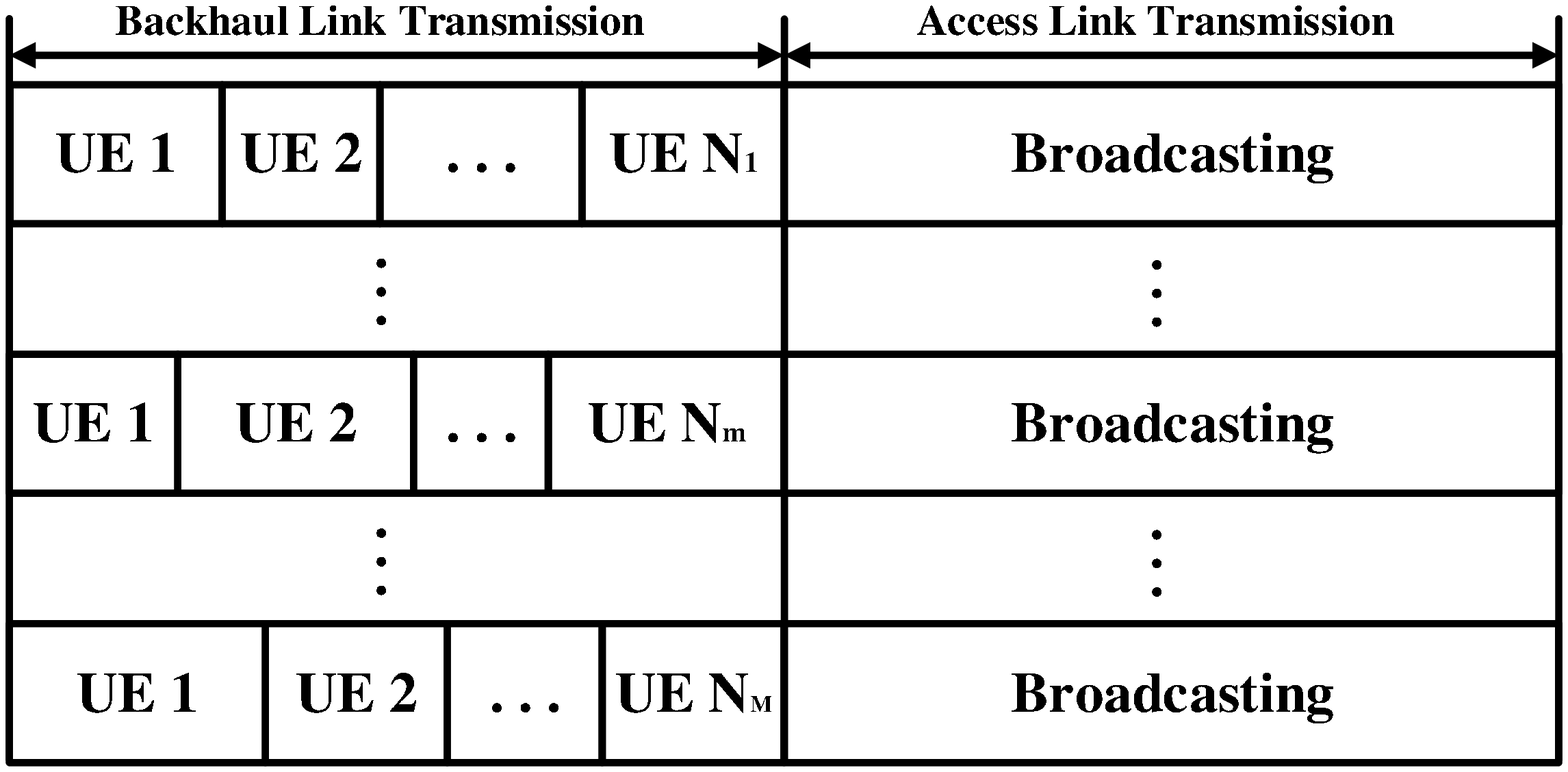}
    \vspace{0.2 cm}
  \caption{An illustration of the backhaul link sharing scheme.}\label{fg:002}
\end{figure}

We consider the downlink transmission of SCN, which consists of an $L$-antenna gateway and $M$ single-antenna ScBSs.
The $m$-th ScBS is associated with $N_m$ single-antenna user equipments (UEs), $m = 1, 2, \ldots, M$.
As shown in Fig. \ref{fg:001}, the ScBSs communicate with the gateway and associated UEs via the MISO-BC backhauls and RF access links, respectively.
The MISO-BC backhauls and RF access links operate in the orthogonal time slots in order to avoid the interference.
Specifically, the gateway uses half portion of the frame for backhauling transmission as shown in Fig. \ref{fg:002}.
The ScBSs use the other half portion of the frame for access-link transmission as shown in Fig. \ref{fg:002}.

The ScBSs operate in the same channel.
Therefore, the inter-cell and intra-cell interference exist among the \mbox{ScBSs}.
We assume that the channel state information (CSI) of \mbox{MISO-BC} backhauls and RF access links is perfectly known at the gateway.
Specifically, the CSI at receivers can be obtained via channel estimation of downlink pilots.
The CSI at transmitters can be obtained via uplink feedback in \mbox{frequency-division} duplex mode or channel reciprocity estimation in the \mbox{time-division} duplex mode \cite{DongJan.2019}.

\subsection{Signal Models}
\subsubsection{Signal Models in Backhauls}
At the gateway, the information-bearing signal for the $m$-th ScBS is defined as $\bm w_m \in \mathbb{C}^{L\times 1}$.
Hence, the transmit signal at the gateway is given as
\begin{equation}\label{eqa:02}
\bm w = \summ \bm w_m.
\end{equation}
The overall transmit covariance matrix is obtained as $\W = \sum\nolimits_{m=1}^M \W_m$, where $\W_m \triangleq \mathds{E}\cb{\bm w_m \bm w^{\hh}_m}$ is the transmit covariance matrix for the $m$-th ScBS.

Let $\h_m \in \mathbb{C}^{L\times 1}$ denote the channel-coefficient vector between the gateway and the $m$-th ScBS (the $m$-th backhaul link).
The Rayleigh fading is considered; therefore, each entry of the $m$-th channel-coefficient vector $\h_m$ follows a circularly-symmetric-complex-Gaussian (CSCG) distribution $\cn\br{0, \Omega_m^{-1}}$.
Here, the value $\Omega_m$ denotes the pathloss of the $m$-th link.

The received signal at the $m$-th ScBS is denoted as
\begin{equation}\label{eqa:03}
y_m = \h_m^{\hh}\bm w + z_m
\end{equation}
where $z_m$ is the additive white Gaussian noise with the CSCG distribution as $\cn\br{0, \sigma^2}$.

Based on the DPCS, the gateway communicates with the ScBSs via the MISO-RF backhauls.
In the DPCS, the information-bearing signals are sequentially encoded.
Let $\bm\pi \triangleq \cb{\pi_1, \pi_2, \ldots, \pi_M}$ denote the encoding order of ScBSs.
Here, the term $\pi_m$ denotes the index of \mbox{information-bearing} signals encoded at the \mbox{$\br{M+1-m}$-th} order.
Therefore, the channel capacity of backhauls is denoted as \cite{WangMay2015}
\begin{equation}\label{eqa:04}
\begin{split}
 & C_{\B}\br{\bm\pi, \cb{\h_m, \W_m}_{\forall m}} \\
=& \cb{
R^{\B}_{\pi_m} \le s \log\frac{{\Psi_{\pi_m}}}{{\Theta_{\pi_m}}},
\st \summ \tr\br{\W_m} \le P^{\max}_{0} }
\end{split}
\end{equation}
where
\begin{equation}\label{eqa:05}
\Psi_{\pi_m} \triangleq {\h_{\pi_m}^{\hh} \sum\limits_{k=1}^m \W_{\pi_k} \h_{\pi_m}  + \sigma^2}
\end{equation}
and
\begin{equation}\label{eqa:06}
\Theta_{\pi_m} \triangleq {\h_{\pi_m}^{\hh} \sum\limits_{k=1}^{m-1} \W_{\pi_k} \h_{\pi_m}  + \sigma^2}
\end{equation}
with $\Theta_{\pi_1} = \sigma^2$.
Here, $P^{\max}_{0}$ is the maximum transmit power of gateway.

\subsubsection{Signal Models in Access Links}
After decoding the information-bearing signals, the $m$-th ScBS broadcasts the information-bearing signals to the associated UEs.
The received signal of the $n$-th UE at the $m$-th ScBS (the $\br{m,n}$-th UE) is denoted as
\begin{equation}\label{eqa:07}
\begin{split}
y_{m,n} =& g_{m,n} \sqrt{v_{m,n}} x_{m,n} + g_{m,n} \sum\limits_{i \neq n} \sqrt{v_{m,i}} x_{m,i} \\
&+ \sum\limits_{j \neq m}\sum\limits_{i = 1}^{N_j} g_{j,n} \sqrt{v_{j, i}} x_{j,i}  + z_{m,n}
\end{split}
\end{equation}
where $g_{m,n}$, $x_{m,n}$ and $v_{m,n}$ are, respectively, the channel coefficient,  information-bearing signal and transmit power for the $\br{m,n}$-th UE.
Here, $g_{m,n}$ follows a CSCG distribution $\cn\br{0, \omega_{m,n}^{-1}}$ with $\omega_{m,n}$ as the pathloss of the $\br{m,n}$-th UE to the $m$-th ScBS.
The term $z_{m,n}$ is the AWGN at the \mbox{$\br{m,n}$-th} UE with CSCG distribution $\cn\br{0, \sigma^2}$.

Based on \eqref{eqa:07}, the received SINR at the $\br{m,n}$-th UE is denoted as
\begin{equation}\label{eqa:08}
\sinr^{\A}_{m,n} = \frac{v_{m,n}\abs{g_{m,n}}^2}{I^{\intra}_{m,n} + I^{\inter}_{m,n} + \sigma^2}
\end{equation}
where $I^{\intra}_{m,n}$ and $I^{\inter}_{m,n}$ are, respectively, defined as
\begin{equation}\label{eqa:09}
I^{\intra}_{m,n} \triangleq \sum\limits_{i \neq n} v_{m,i}\abs{g_{m,n} }^2
\end{equation}
and
\begin{equation}\label{eqa:10}
I^{\inter}_{m,n} \triangleq \sum\limits_{j \neq m}\sum\limits_{i = 1}^{N_j} v_{j, i} \abs{g_{j,n}}^2.
\end{equation}

%

\subsection{Problem Formulation}
Since the $m$-th backhaul link is shared by the $N_m$ UEs of the $m$-th ScBS, the flow-conservation constraint of the $m$-th backhaul is denoted as
\begin{equation}\label{eqa:11}
R^{\B}_m \ge \sumn R^{\A}_{m,n}
\end{equation}
where $R^{\A}_{m,n} = \log\br{1+\sinr^{\A}_{m,n}}$.

Our objective is to minimize the system transmit power of the gateway and ScBSs via
the joint design of precoding covariance matrices $\cb{\W_m}_{\forall m}$ and proportional ratios $\cb{\phi_m}_{\forall m}$ of MISO-BC backhauls, the power-control variables $\cb{v_{m,n}}_{\forall m, n}$ of access links.
Let $\Y$ denote the set of optimization variables as $\Y \triangleq \cb{\W_m, \phi_m, v_{m,n}}_{\forall m, n}$, where ${\phi_m}$ is the ratio of rate that is shared by the $m$-th backhaul link with $\sum\nolimits_{m=1}^M \phi_m = 1$.
In order to minimize the system transmit power, we formulated the JPPc problem  as
\begin{subequations}\label{eqa:12}
\begin{align}
\min\limits_{\Y} \;&  \summ \tr(\W_m) + \summ\sumn{v_{m,n}}  \label{eqa:12a}\\
\st & \summ \tr\br{\W_m} \le P_{0}^{\max} \label{eqa:12b}\\
& R_1^{\B}: \ldots: R_M^{\B} = \phi_1: \ldots: \phi_M \label{eqa:12c}\\
& R^{\B}_m \ge \sumn R^{\A}_{m,n}, \forall m \label{eqa:12d}\\
& \sumn v_{m,n} \le P_m^{\max}, \forall m \label{eqa:12e} \\
& R^{\A}_{m,n} \ge R^{\req}_{m,n}, \forall m, n \label{eqa:12f}
\end{align}
\end{subequations}
where $P_m^{\max}$ and $R^{\req}_{m,n}$ are respectively the maximum transmit power of the $m$-th ScBS and the communication QoS requirement of the $\br{m,n}$-th UE.

\section{Centralized Optimal Solution}
Since the JPPc problem \eqref{eqa:12} contains the \mbox{non-convex} proportional-rate constraints in \eqref{eqa:12c} and the \mbox{non-convex} flow-conservation constraints in \eqref{eqa:12d}, the optimization problem is challenging to handle with standard optimization tools.
Therefore, we are motivated to investigate the structure of JPPc problem \eqref{eqa:12} such that an optimal solution is obtained.

Based on the structure of JPPc problem \eqref{eqa:12}, we propose a two-stage optimization framework.
In the optimization framework, the JPPc problem \eqref{eqa:12} is decoupled into access-link power-control subproblem and backhaul precoding subproblem.
After several algebraic manipulations, we obtain the optimal power-control variables via the off-the-shelf toolbox, e.g., CVX \cite{Grant2014}.
Solving the backhaul precoding subproblem, we obtain the optimal closed-form precoding vectors for the gateway.

\subsection{Optimal Proportional Ratios for MISO-BC Backhauls}
Before analyzing the optimal proportional ratios, we first introduce a proposition as follows.
\begin{proposition}\label{pr:01}
Ignoring the proportional-rate constraints in \eqref{eqa:12c}, the constraints in \eqref{eqa:12d} and \eqref{eqa:12f} are active when the problem \eqref{eqa:12} is optimally solved.
\end{proposition}
\begin{IEEEproof}
See Appendix \ref{apdx:01}.
\end{IEEEproof}

When the constraints \eqref{eqa:12d} and \eqref{eqa:12f} are active, the required rate for the $m$-th backhaul is obtained as
\begin{equation}\label{eqa:13}
R_m^{\B} = R_m^{\req} \triangleq \sum\limits_{n=1}^{N_m} R_{m,n}^{\req}
\end{equation}
with $m = 1, 2, \ldots, M$.

Substituting \eqref{eqa:13} into \eqref{eqa:12c}, we obtain the optimal proportional ratios $\cb{\phi_m^*}_{\forall m}$ as
\begin{equation}\label{eqa:14}
\phi_m^* = \frac{R_{m,n}^{\req}}{\sum\limits_{n=1}^{N_m} R_{m,n}^{\req}}
\end{equation}
otherwise, more transmit power is required to the MISO-BC backhauls.

Based on Proposition \ref{pr:01} and performing several algebraic manipulations, we equivalently decouple the JPPc problem \eqref{eqa:12} into access-link power-control subproblem and backhaul precoding subproblem as
\begin{subequations}\label{eqa:16}
\begin{align}
\min\limits_{\cb{v_{m,n}}_{\forall m,n}} \;& \summ\sumn{v_{m,n}} \label{eqa:16a}\\
\st & \sumn v_{m,n} \le P_m^{\max}, \forall m \label{eqa:16b} \\\
    & \sinr_{m,n} = \Gamma^{\req}_{m,n}, \forall m, n \label{eqa:16c}
\end{align}
\end{subequations}
and
\begin{subequations}\label{eqa:15}
\begin{align}
\min\limits_{\cb{\W_m}_{\forall m}} \;&  \summ \tr\br{\W_m} \label{eqa:15a}\\
\st & \summ \tr\br{\W_m} \le P_{0}^{\max} \label{eqa:15b}\\
& R_m^{\B} = R_m^{\req}, \forall m \label{eqa:15c}
\end{align}
\end{subequations}
where $\Gamma_{m,n}^{\req} = \exp\br{R_{m,n}^{\req}} - 1$.

\subsection{Optimal Power control for Access Links}
In order to solve the access-link power-control subproblem \eqref{eqa:16}, we obtain a set of equivalent constraints to \eqref{eqa:16c} as
\begin{multline}\label{eqa:17}
\frac{v_{m,n}\abs{g_{m,n}}^2}{\Gamma_{m,n}^{\req}} \\
= \sum\limits_{i \neq n} v_{m,i}\abs{g_{m,n} }^2 + \sum\limits_{j \neq m}\sum\limits_{i = 1}^{N_j} v_{j, i} \abs{g_{j,n}}^2 + \sigma^2.
\end{multline}

Substituting \eqref{eqa:17} into \eqref{eqa:16}, we obtain a convex version of access-link power-control subproblem as
\begin{subequations}\label{eqa:18}
\begin{align}
\min\limits_{\cb{v_{m,n}}_{\forall m,n}} \;& \summ\sumn{v_{m,n}} \label{eqa:18a}\\
\st & \sumn v_{m,n} \le P_m^{\max} \mbox{ and } \eqref{eqa:17}, \forall m. \label{eqa:18b}
\end{align}
\end{subequations}

Since the optimization problem \eqref{eqa:18} is convex, the optimal power-control variables $\cb{v_{m,n}}_{\forall m,n}$ can be obtained via second-order cone programming or semidefinite programming.

\subsection{Optimal Precoding Vectors for MISO-BC Backhauls}
We propose a new method to obtain the closed-form optimal solution.
Based on the uplink-downlink duality, we obtain a set of equivalent convex constraints to \eqref{eqa:15c}.

The received signal in the dual uplink channel of \eqref{eqa:03} is obtained as
\begin{equation}\label{eqa:19}
\overline{\bm y} = \summ \h_m \sqrt{\overline{w}_m} x_m + \overline{\bm z}
\end{equation}
where $x_m$, ${\overline{w}_m}$ and $\overline{\bm z}$ are, respectively, dual signal of the $m$-th ScBS, the power of dual signal and additive white Gaussian noise at the gateway.
Here, the AWGN at the gateway follows CSCG distribution with mean zero and covariance matrix $\sigma^2\bm I$.

Since the downlink channel capacity with the encoding order $\pi$ in \eqref{eqa:03} is equal to the dual uplink channel capacity with the inverse decoding order  as \cite{WangMay2015, VishwanathOct.2003}
\begin{equation}\label{eqa:20}
\begin{split}
 & C_{\D}\br{{\bm\pi}, \cb{\h_m, \overline w_m}_{\forall m}} \\
=& \cb{R^{\D}_{\pi_m} \le  \log\frac{\abs{\overline{\bm\Psi}_{\pi_m}}}{\abs{\overline{\bm\Theta}_{\pi_m}}}, \st \summ \overline w_m \le P_{0}^{\max} }
\end{split}
\end{equation}
where  $\overline{\bm\Psi}_{\pi_m}$ and $\overline{\bm\Theta}_{\pi_m}$ are, respectively, defined as
\begin{equation}\label{eqa:21}
\overline{\bm\Psi}_{\pi_m} \triangleq \sum\limits_{k=m}^M \h_{\pi_k} \overline w_{\pi_k} \h_{\pi_k}^{\hh}  + \sigma^2\bm I
\end{equation}
and
\begin{equation}\label{eqa:22}
\overline{\bm\Theta}_{\pi_m} \triangleq  \sum\limits_{k=m+1}^{M} \h_{\pi_k} \overline w_{\pi_k} \h_{\pi_k}^{\hh}  + \sigma^2\bm I
\end{equation}
with $\overline{\bm\Theta}_{\pi_M} = \sigma^2\bm I$.

Moreover, the injection between $\overline w_{\pi_m}$ and $\W_{\pi_m}$ is defined as \cite{VishwanathOct.2003}
\begin{equation}  \label{eqa:23}
\W_{\pi_m}
= \overline{\bm\Theta}_{\pi_m}^{-\frac{1}{2}} \bm u_{\pi_m}
\Theta_{\pi_m}
\overline w_{\pi_m}
\bm u_{\pi_m}^{\hh} \overline{\bm\Theta}_{\pi_m}^{-\frac{1}{2}}
\end{equation}
{\color{black}
where the vector $\bm u_{\pi_m}$ is obtained via singular-value decomposition as  $\overline{\bm\Theta}_{\pi_m}^{-\frac{1}{2}} \h_{\pi_m} \Theta_{\pi_m}^{-\frac{1}{2}} = \bm u_{\pi_m} \lambda_{\pi_m}$ with $\bm u_{\pi_m}^{\hh}\bm u_{\pi_m} = 1$ and eigen-value $\lambda_{\pi_m}$.}


Based on the duality of uplink and downlink channels, we obtain
\begin{equation}\label{eqa:24}
R^{\B}_{\pi_{m}} = R^{\D}_{\pi_m} =  {\log\abs{\overline{\bm\Psi}_{\pi_m}} -  \log\abs{\overline{\bm\Theta}_{\pi_m}}}, \forall m
\end{equation}
with $\sum\nolimits_{m=1}^{M} w_{\pi_m} = \sum\nolimits_{m=1}^{M} \tr\br{\W_{\pi_m}}$.


Dropping the power constraint in \eqref{eqa:15b} and replacing the backhaul rate $R_{\pi_m}^{\B}$ in \eqref{eqa:15c} with \eqref{eqa:24}, the optimization problem \eqref{eqa:15} is transformed as
\begin{subequations}\label{eqa:25}
\begin{align}
\min\limits_{\cb{\overline w_{\pi_m}}_{\forall m}} \;& \summ \overline w_{\pi_m} \label{eqa:25a}\\
\st & \log\abs{\overline{\bm\Psi}_{\pi_m}} - \log\abs{\overline{\bm\Theta}_{\pi_m}} =  R_{\pi_m}^{\req}, \forall m. \label{eqa:25b}
\end{align}
\end{subequations}

The optimization problem \eqref{eqa:25} is equivalent to problem \eqref{eqa:15} if and only if the optimal value of \eqref{eqa:25} is less than or equal to $P_0^{\max}$.
Based on \eqref{eqa:20}--\eqref{eqa:22}, we observe that the data rate of the $m$-th backhaul increases with $\overline w_{\pi_m}$.
Moreover, the data rate of the $k$-th backhaul decreases with the power of dual signal $\overline w_{\pi_m}$ when $k < m$.
Therefore, setting the constraints in \eqref{eqa:25b} active, we obtain the optimal power of dual signal $\cb{\overline w^*_{\pi_m}}_{\forall m}$ as
\begin{equation}\label{eqa:26}
\overline{w}_{\pi_m} = \frac{\exp\br{R_{\pi_m}^{\req}} - 1}{\h^{\hh}_{\pi_m}\overline{\bm\Theta}_{\pi_m}^{-1}\h_{\pi_m}}.
\end{equation}

Substituting \eqref{eqa:26} into \eqref{eqa:23} and performing some algebraic manipulations, we obtain the closed-form optimal downlink precoding vectors $\cb{\w_{\pi_m}}_{\forall m}$ as
\begin{equation}\label{eqa:27}
\w_{\pi_m} = \frac{\exp\br{R_{\pi_m}^{\req}}-1}{\h^{\hh}_{\pi_m}\overline{\bm\Theta}_{\pi_m}^{-1}\h_{\pi_m}}
\frac{\overline{\bm\Theta}^{-1}_{\pi_m}\h_{\pi_m}}
{\norm{\overline{\bm\Theta}^{-\frac{1}{2}}_{\pi_m}\h_{\pi_m}\Theta_{\pi_m}^{-\frac{1}{2}}}}.
\end{equation}

After obtaining the CSI of backhauls and access links, the gateway can obtain the optimal precoding vectors in two steps: 1) calculating the set of optimal power of dual signals for dual uplink channel based on \eqref{eqa:26}; and 2) calculating the optimal precoding vectors for downlink channel based on \eqref{eqa:27}.
Based on \eqref{eqa:14}, \eqref{eqa:18} and \eqref{eqa:27}, we can obtain the optimal proportional ratios, access-link power-control variables and backhaul precoding vectors.

\section{Simulation Results}
We use numerical results to verify our proposed algorithm.
In order to illustrate the performance improvement, we also include a benchmark scheme where the backhauls use \mbox{zero-forcing} beamforming.
The pathloss equations for the backhauls and access links are, respectively, given as \cite{tr38901}
\begin{equation}
\Omega_m = 32.4 + 20\log_{10}\br{ f_c } + 31.9\log_{10}\br{ D_m } \mbox{ dB}
\end{equation}
and
\begin{equation}
\omega_{m,n} = 17.3 + 24.9\log_{10}\br{ f_c } + 38.3\log_{10}\br{ d_{m,n} } \mbox{ dB}
\end{equation}
where $D_m$ and $d_{m,n}$ are the distances of $m$-th backhaul and the $\br{m,n}$-th access link, respectively.
The power of AWGN is $-107$ dBm\footnote{This value is obtained when power spectrum density of AWGN is $-174$ dBm/Hz,  the noise figures of gateway and UE are $5$ dB, the noise figure of ScBS is $9$ dB, and the
bandwidth of system is $200$ KHz.}.
The gateway is equipped with eight antennas.
There are four ScBSs, and each ScBS is associated with one UEs.
The maximum transmit powers of the gateway and ScBSs are $30$ dBm and $23$ dBm.

We define the system outage event as either the backhauls or the access links are in outage.
When the system outage event happens, the gateway and ScBSs will not transmit any information.

\begin{figure}[htb]
\centering
\includegraphics[width=3 in]{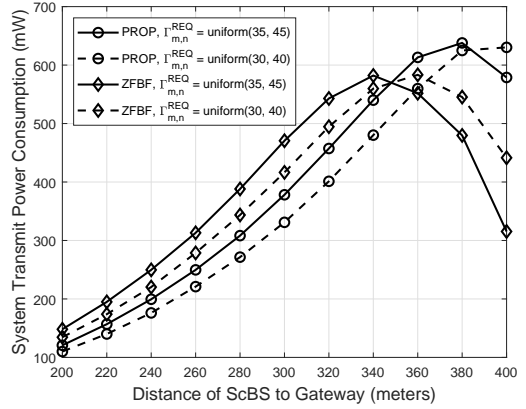}
\caption{The variation of system power consumption over the distance of backhauls.}\label{fg:003}
\end{figure}
\vspace{-0.6 cm}
\begin{figure}[htb]
\centering
\includegraphics[width=3 in]{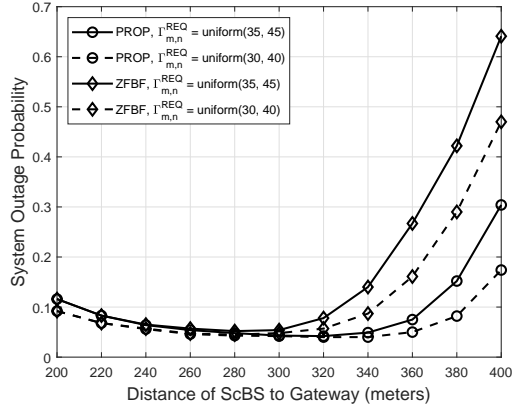}
\caption{The variation of system outage over the distance of backhauls.}\label{fg:004}
\end{figure}

Figure \ref{fg:003} shows the variation of the system transmit power over the distance of backhauls.
We observe that the system transmit power increases with the distance of backhauls.
After reaching the peak value, the system transmit power starts to decrease.
For example, the system transmit power decreases after the distance of backhauls is greater that $380$ meters.
These two observations can be explained as follows.
As the distance of backhauls increases, the required power of the backhauls to deliver a certain amount of information increases.
Since the gateway has a limitation on the maximum transmit power, the system outage probability increases as shown in Fig. \ref{fg:004}.
When a certain threshold is surpassed, the SCN has a high probability to be in an outage event.
In other words, the SCN will be silent with a high probability.
Therefore, the system transmit power decreases.

Figure \ref{fg:004} illustrates the variation of the system outage probability over the distance of backhauls.
We observe that the outage probability of the SCN also increases when the distances of fronthaul links decrease.
This observation can be explained as follows.
When the distances of fronthaul links decrease, the outage probability of access links increases due to the increasing interference among the access links.

Based on Fig. \ref{fg:003} and Fig. \ref{fg:004}, we observe that our proposed scheme outperforms the ZFBF scheme which is used \cite{MostafaDec.2018}.
When the required SINR of UEs are uniformly drawn from the ranges $\br{35, 45}$ and $\br{30, 40}$, our proposed scheme can respectively reduce at most $20.5$\% and $21.0$\% of system transmit power when compared with the ZFBF scheme.
When the required SINR of UEs are uniformly drawn from $\br{35, 45}$, our proposed scheme can reduce the outage probability by $19.23$\% compared with the ZFBF scheme.
Moreover, our proposed scheme can reduce the outage probability by $11.11$\% compared with the ZFBF scheme when the required SINR of UEs are uniformly drawn from $\br{30, 40}$.

\vspace{-0.3 cm}
\section{Conclusions}
We investigated the structure of the formulated JPPc problem in order to optimally minimize the system transmit power with proportional-rate constraints for MISO-BC backhauls.
Based on the structure of the JPPc problem, we obtained the optimal ratios for the backhauls such that we can separate the precoding vectors and power-control variables into two subproblems: access-link power-control subproblem and backhaul precoding subproblem.
The optimal power-control variables can be obtained via the standard convex optimization toolbox.
Leveraging the information-theoretical uplink-downlink duality, we obtain the closed-form expression of the precoding vectors.
Simulation results are used to show the performance improvement over the benchmark scheme.

\vspace{-0.3 cm}
\appendices
\section{Proof of Proposition \ref{pr:01}}\label{apdx:01}
Suppose that the optimal power-control variables $\cb{v_{m,n}}_{\forall m, n}$ do not guarantee that constraints in \eqref{eqa:12f} are inactive.
Without loss of generality, we assume that the $\br{m,n}$-th constraint in \eqref{eqa:12f} is inactive.
We can obtain a lower transmit power $\tilde v_{m,n} \le v_{m,n}$ such that the $\br{m,n}$-th constraint in \eqref{eqa:12f} is active.
Using the transmit power $\tilde v_{m,n}$, we observe that the remaining constraints in \eqref{eqa:12f} are still satisfied.
Besides, we have
\begin{equation}\label{eqa:apdx:01}
 \sum\limits_{m=1}^M\sum\limits_{n=1}^{N_m} v_{m,n}
\ge  \sum\limits_{m=1}^M\sum\limits_{n=1}^{N_m} v_{m,n} + \tilde v_{m,n} - v_{m,n}.
\end{equation}

Based on \eqref{eqa:apdx:01}, the power-control variables $\cb{v_{m,n}}_{\forall m, n}$ is not optimal.
Therefore, we obtain the contradiction.
We conclude that the optimal power-control variables $\cb{v_{m,n}}_{\forall m, n}$ guarantee that constraints in \eqref{eqa:12f} are active.

When the constraint in \eqref{eqa:12c} are ignored, we can use the similar arguments to prove that the constraints in \eqref{eqa:12d} are active.

\bibliographystyle{IEEEtran}
\bibliography{dyj_bib}
\end{document}